\documentclass[preprint2]{aastex702}

\usepackage{amsmath}
\usepackage{hyperref}
\usepackage{graphicx}
\usepackage{CJK}

\shorttitle{CMR Kinematics}
\shortauthors{Kong et al.}

\begin{document}

\title{Filament Formation via Collision-induced Magnetic Reconnection -- Kinematic Features}

\author[0000-0002-8469-2029]{Shuo Kong}
\affiliation{Steward Observatory, University of Arizona, Tucson, AZ 85719, USA}
\email[show]{shuokong@arizona.edu}

\author[0000-0002-7082-0587]{Griselda Arroyo-Ch\'avez}
\affiliation{Steward Observatory, University of Arizona, Tucson, AZ 85719, USA}
\email{arroyochavezg@arizona.edu}

\author[0000-0002-8351-3877]{Volker Ossenkopf-Okada}
\affiliation{I.~Physikalisches Institut, Universit\"at zu K\"oln, Z\"ulpicher Str. 77, 50937 K\"oln, Germany}
\affiliation{Cluster of Excellence ``Our Dynamic Universe'' (DYNAVERSE)}
\email{ossk@ph1.uni-koeln.de} 

\author[0000-0001-5653-7817]{H\'ector G. Arce}
\affiliation{Department of Astronomy, Yale University, New Haven, CT 06511, USA}
\email{hector.arce@yale.edu}

\author[0000-0002-0560-3172]{Ralf S. Klessen}
\affiliation{Universit\"{a}t Heidelberg, Zentrum f\"{u}r Astronomie, Institut f\"{u}r Theoretische Astrophysik, Albert-Ueberle-Str. 2, 69120 Heidelberg, Germany}
\affiliation{Universit\"{a}t Heidelberg, Interdisziplin\"{a}res Zentrum f\"{u}r Wissenschaftliches Rechnen, Im Neuenheimer Feld 205, 69120 Heidelberg, Germany}
\email{klessen@uni-heidelberg.de}

\author[0000-0001-6216-8931]{Duo Xu}
\affiliation{Canadian Institute for Theoretical Astrophysics, University of Toronto, 60 St. George Street, Toronto, ON M5S 3H8, Canada}
\email{xuduo@cita.utoronto.ca}

\author[0009-0009-1088-9062]{Szu-Ting Chen}
\affiliation{Steward Observatory, University of Arizona, Tucson, AZ 85719, USA}
\email{szutingchen@arizona.edu}

\begin{abstract}
The Collision-induced Magnetic Reconnection (CMR) mechanism
has recently been proposed as a filament formation model
that emphasizes the active role of magnetic fields.
To enable better observational tests with data from modern telescopes,
we provide more specific observable predictions for filaments
formed via CMR.
Two types of CMR kinematics are revealed in
position-velocity (PV) diagrams.
First, due to the
nature of CMR, the midplane contains a {\it converging motion}
close to the filament and two {\it diverging motions} farther from
the filament. They exhibit a blueshift, redshift, blueshift, redshift
(BRBR) velocity pattern along the line-of-sight when the collision
midplane is inclined, giving rise to
a special pattern in the PV-diagram. Second,
the longitudinal PV-diagram along the filament spine exhibits
a velocity oscillating pattern, which is currently attributed to
the magnetic transport of gas clumps
in the {\it converging motion}.
The first type of pattern involves emission primarily
outside the filament, which can be confused with the environment.
The second type, however, mainly emerges from
gas inside the filament, thus more robust for observational
tests. Both the integral-shaped filament and the Stick filament
in the Orion A cloud show the velocity oscillation in PV-diagrams,
although the spatial oscillation frequency and their velocity spread
differ.
\end{abstract}

\begin{keywords}
    {Star Formation, Filaments}
\end{keywords}


\section{Introduction}\label{sec:intro}

Molecular clouds play a fundamental role in the process
of star formation. Far-infrared surveys have demonstrated
that these clouds are predominantly filamentary in nature
\citep{2010A&A...518L.100M,2011A&A...529L...6A,Andre+2014,Hacar+2018}.
Consequently, elucidating the formation and evolution of filamentary
structures constitutes a critical step toward a comprehensive
understanding of star formation. Magnetic fields are
also known to play a significant role 
\citep{Mouschovias.Spitzer1976,2012ARA&A..50...29C,Pudritz+2014,Planck+2016XXXV,Hennebelle.Inutsuka2019}
in the dynamics of the interstellar medium (ISM),
and their influence
on filament formation is essential for developing a
complete physical framework of the star formation process
\citep{2000MNRAS.311...85F}.

Over the years, a variety of filament formation mechanisms
have been proposed across different physical scales.
Early theoretical work by \citet{Ostriker1964} and 
\citet{Inutsuka.Miyama1992} established the foundations for 
the gravitational stability of filamentary structures,
particularly in the context of isothermal cylinders.
More recently, different formation scenarios have emphasized
specific properties of the interstellar medium. 
These include the formation of filaments by MHD shocks with turbulence 
\citep{Padoan+2001,Federrath2016,2020ApJ...891..168W,2020MNRAS.494.3675C},
the development of filamentary structures through thermal instability
\citep{Audit.Hennebelle2005,Vazquez-Seamdeni+2006,Saury+2014},
and gravity dominated formation and evolution mechanisms 
\citep{Hartmann.Burkert2007,Banerjee+2009,Gomez.Vazquez-Semadeni2014,Vazquez-Semadeni+2019}.
On larger scales, galactic shear by differential rotation
has also been proposed as a mechanism for the formation of
elongated structures, including the so-called ``bones'' of the Milky Way 
\citep{Goodman+2014,Zucker+2015,2020MNRAS.492.1594S}.
A detailed discussion of the different proposed mechanisms for
filament formation can be found in the recent review
by \citet{2023ASPC..534..153H}.

\begin{figure*}[htb!]
\centering
\includegraphics[width=0.95\textwidth]{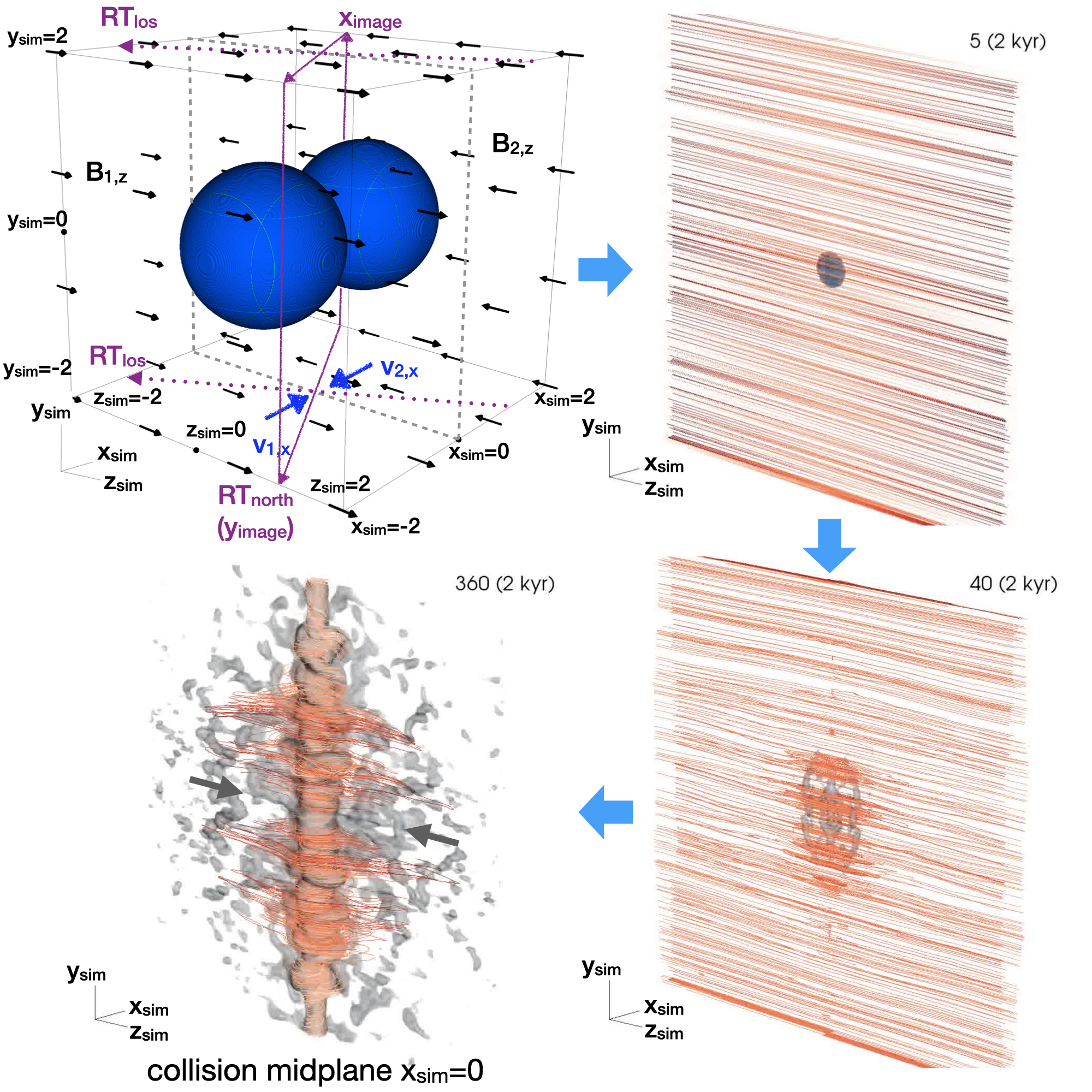}
\caption{
Illustration of the CMR process.
The top-left panel shows a 3D view of the initial condition.
The simulation coordinate system ($x_{\rm sim}$, $y_{\rm sim}$, $z_{\rm sim}$) is in units of pc.
Two spherical clumps (each with a radius of 0.9 pc)
collide along $x_{\rm sim}$ with velocities
$v_{\rm 1,x}$ and $v_{\rm 2,x}$.
Magnetic fields (black vectors)
reverse across the $x_{\rm sim}=0$ plane (gray dashed lines).
The purple dotted arrows show the RT-model line-of-sight (RT$_{\rm los}$).
The purple solid lines show the image plane
perpendicular to the line-of-sight. The image coordinate system
($x_{\rm image}$, $y_{\rm image}$) are shown as purple arrows.
$y_{\rm image}$ corresponds to the north direction of the projection (RT$_{\rm north}$).
The remaining 3 panels show the filament formation process 
as a function of time (evolving clockwise).
They show magnetic fields (streamlines) and 
dense gas (volume-rendering) across the collision midplane ($x_{\rm sim}=0$).
The two gray arrows in the lower-left panel 
indicate the {\it converging motion} of dense gas pieces.
They are magnetically transported to the filament.
At the end, the filament has a length of $\sim$2 pc.
The full movie is available \href{https://doi.org/10.7910/DVN/CXHWRR}{here}
with a file name {\tt CMRfield.mpg}.
\label{fig:model}}
\end{figure*}

Motivated by observations of the Stick filament in the Orion A molecular
cloud, \citet[][hereafter K21]{2021ApJ...906...80K}
proposed the Collision-induced Magnetic Reconnection
(CMR) mechanism as a novel pathway for filament formation.
In this scenario, a collision between two parcels of magnetized gas
carrying oppositely-oriented magnetic field triggers
the formation of a filamentary structure.
While we refer readers to K21 and 
\citet{2023ApJS..265...58K}
for a detailed description of the simulation,
the key distinction between
CMR and other filament formation models is the role of
magnetic fields. In CMR, the formation of the filament
is purely due to magnetic fields and does not require gravity
(although gravity  plays an important role later, as the cloud
grows in mass). The key physics in CMR is the following.
As the oppositely directed field lines are pushed together,
magnetic reconnection is triggered and field loops encircle
the compression layer (a pancake if the initial colliding parcels
are spheres). The field loop exerts strong 
magnetic tension and redirects and concentrates gas
into a dense filament, which is at the location of the symmetric axis.
While magnetic fields are 
either secondary to gravity/turbulence
or play a largely {\it passive} role in most
models, they play a dominating and {\it active} role in CMR.
In turn, CMR reveals a fundamentally different dynamical role
of magnetic fields in the ISM.
Figure \ref{fig:model} gives a detailed illustration
for CMR (see \S\ref{subsec:sim}).

So far, this mechanism has been tested in the case of the Stick filament
\citep[see also][]{2025ApJS..280...25Z}.
A natural question is whether the same mechanism happens more 
in the ISM. There is a growing number of detections of
magnetic field reversals in the interstellar medium 
\citep{1997ApJS..111..245H,2018A&A...614A.100T,2018ApJS..234...11H,2022A&A...660A..97T,2022ApJ...940...75D,2026ApJ...997..304B}. In principle,
any colliding flows from two sides of a reversal
can trigger CMR. The question is, what are some observational
features that can decisively distinguish CMR-filaments from others.
Meanwhile, with the growing observational power of modern facilities,
interstellar filaments are revealing increasingly distinct properties,
highlighting the need to test details of filament formation models
in order to better distinguish them. 

In principle, one can identify clear observational cases
in which two gas clouds collide with reversed magnetic
fields. However, it is challenging to determine the physical
conditions that precede filament formation, even if
the filament indeed forms via CMR. A more practical
approach, therefore, is to identify stable observational
signatures that are characteristic of the CMR mechanism.
In practice, there is probably no truly single decisive
feature for observational test. 
Using the CMR model, K21 successfully reproduced several
key properties of the Stick filament: morphological
features (ring- and/or fork-like structures), kinematic
features (a two-velocity pattern in transverse position-velocity
diagrams), and dynamical features (pressure-confined
cores sustained by surface magnetic pressure). In this
paper, we identify
two {\it additional kinematic features} that
can serve as observational diagnostics of CMR. These
features manifest in position-velocity (PV) diagrams,
which are commonly used in the analysis of molecular
line observations of molecular clouds. The features
arise from the distinctive kinematic patterns produced
by CMR, and their uniqueness will be evaluated through
comparisons with other filament formation models in the future.

In the following section \S\ref{sec:data},
we introduce the data used for the analysis.
In \S\ref{sec:results}, we show in detail the kinematic
features in PV-diagrams and their origin.
The features can be used as observational diagnostics
for CMR-filaments.
In \S\ref{sec:discus}, we discuss the features in
the larger Orion A cloud. 
In \S\ref{sec:sum}, we summarize and conclude the paper.

\section{Data}\label{sec:data}

\subsection{Simulation Data}\label{subsec:sim}

We focus on the fiducial simulation \mbox{MRCOL} in K21
performed with the grid-based MHD code \textsc{Athena++}
\citep{2020ApJS..249....4S}.
The simulation setup and physical conditions have
been described in detail in K21 (also see \citealt{2023ApJS..265...58K},
Figure 1 and Table 1). Here we briefly repeat the
key parameters. The colliding clumps both have a density of
$n_{\rm H_2}=420$ cm$^{-3}$. The ambient gas density is a factor
of 10 smaller. The computation domain has an isothermal
temperature of 15 K with a uniform field strength of 10 $\mu$G
(reverses at the collision midplane). The plasma-$\beta$ is
thus 0.033, i.e., magnetically dominated, which is why 
in K21 the same clump collision but without reverse fields
results in no dense gas formation at all (see their Figure 12).
Each clump moves at a speed of 1 km s$^{-1}$ toward the other.
There is also a 0.25 km s$^{-1}$ shear velocity
so that the collision is at an oblique angle with
respect to the magnetic field discontinuity.
Later in the Discussion section where we introduce
the \mbox{test\_Bpad} simulation and another simulation
from \citet{2024ApJ...975...97K}, the shear velocity is
removed so that the collision is head-on. In \mbox{MRCOL}, the isothermal
sound speed is 0.29 km s$^{-1}$ and
Alfv\'en speed is 0.64 km s$^{-1}$.
To alleviate boundary artifacts, 
we expand the computation domain by a factor of 2
and re-run the simulation (hereafter \mbox{MRCOLe2}).
All the other simulation parameters
remain the same as \mbox{MRCOL}. 

Figure \ref{fig:model} presents a 3D illustration of
the simulation and the evolution of the filament.
Hereafter, the simulation coordinate system is
denoted as ($x_{\rm sim}$, $y_{\rm sim}$, $z_{\rm sim}$).
The top-left panel shows the initial condition, in which two
spherical clumps collide at the interface between oppositely
directed magnetic fields at $x_{\rm sim}=0$, the collision midplane.
The remaining three panels illustrate the subsequent evolution
of gas (volume-rendering) and magnetic fields (streamlines).
With the onset of the collision,
gas is rapidly compressed at the midplane.
By $t=10$ kyr,
a round dense sheet (hereafter the {\it pancake})
emerges in the midplane, as shown by the top-right panel.
By $t=80$ kyr, this {\it pancake} has developed pronounced
density substructure, while magnetic-field loops begin
to emerge around its edges (bottom-right panel).
As discussed by K21 (Section 3.6),
these loops result from magnetic reconnection
at the two edges of the {\it pancake}. The resulting
field loops squeeze the dense layer toward $y_{\rm sim}$
by transporting dense gas from the surrounding {\it pancake}
into the central filament (bottom-left panel).
We refer to this process as {\it magnetic transport}.
Meanwhile, the reconnected
field lines extending away from the {\it pancake} form
half-loops that pull gas outward from the dense layer.
By $t=720$ kyr, the filament has become the dominant
dense structure and is confined by magnetic pressure
associated with the surrounding helical/toroidal magnetic field.

This evolution can be viewed as a sequential reduction
in dimensionality. The initial three-dimensional clumps
are first compressed into a two-dimensional sheet, the
{\it pancake}, which is subsequently squeezed into a
one-dimensional filament. As the clump-clump collision
persists, dense gas continues to accumulate at the midplane
and is transported toward the filament by newly formed
magnetic-field loops. Thus, the growth of the filament
is sustained for as long as the collision continues.
The process eventually terminates when the collision
loses its driving energy or when longitudinal gravitational
collapse becomes important. For a filament of scale
$\sim2$ pc, \citet{2022MNRAS.517.4679K} showed that
this final stage can lead to collapse into a cluster-forming
core, completing the dimensional sequence from 3D to
2D to 1D and ultimately to 0D.

A movie that includes the snapshots is available on
\href{https://doi.org/10.7910/DVN/CXHWRR}{Dataverse}
(file name {\tt CMRfield.mpg}).
The movie shows more details of the CMR process.
For example, we see both gas moving toward the filament and
away from the filament. Hereafter, we specifically refer to
the motion toward the filament as the {\it converging motion},
and the motion away from the filament as the {\it diverging motion}
(also see Figure \ref{fig:transpv}).
They are both caused by the magnetic tension
of the field loops.
By the end of the simulation,
the filament is encircled by toroidal/helical
fields \citep[see Figure 2 in][]{2023ApJS..265...58K},
which acts as a shield, preventing free gas collapse.
The star formation rate in such a cloud is reduced compared to
other freely collapsing clouds \citep{2022MNRAS.517.4679K}.

\subsection{Synthetic Observation}\label{subsec:so}

\begin{figure*}[htb!]
\centering
\includegraphics[width=0.9\textwidth]{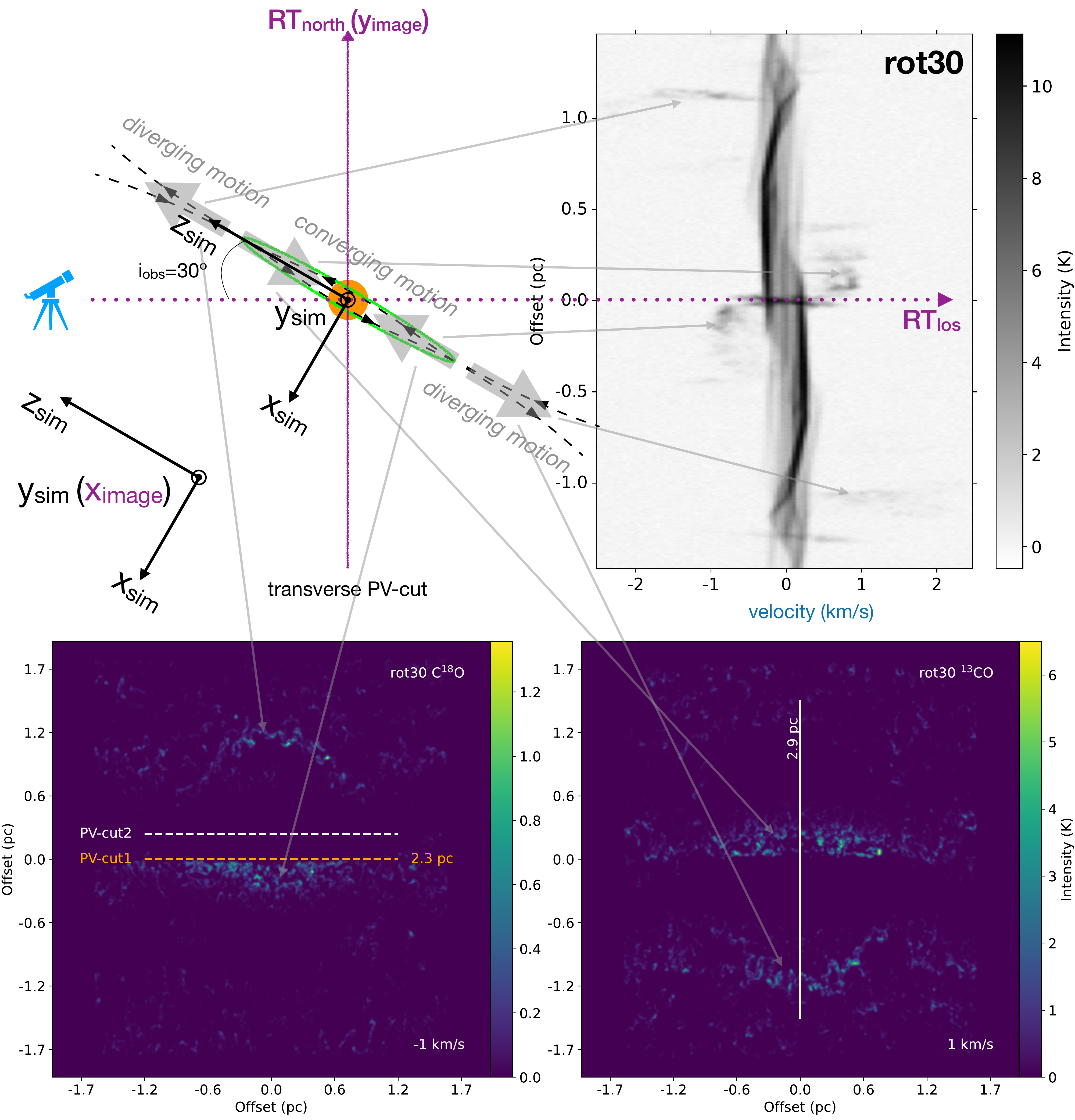}
\caption{
{\bf Top-left:} 
Illustration of the RT setup, following
the convention of Figure \ref{fig:model}.
The simulation coordinate system ($x_{\rm sim}$, $y_{\rm sim}$, $z_{\rm sim}$)
is shown as black arrows. 
Here, we are viewing along the negative $y_{\rm sim}$ direction,
and the slice passes through the origin.
In this view,
we see the cross-section of the filament (orange circle).
The black dashed curves show the field lines,
including the two half-loops at two ends
and the one full-loop in the middle.
The green ellipse shows the imaginary compression {\it pancake}
which is circled by the full-loop.
The thick gray arrows indicate the {\it converging motion} and the {\it diverging motion}.
For the RT-model (purple),
we show the {\tt rot30} projection as we ``observe'' the filament
from left to right.
In this view, $y_{\rm image}$ aligns with RT$_{\rm north}$,
and $x_{\rm image}$ aligns with $y_{\rm sim}$.
{\bf Top-right:} Transverse PV-diagram for $^{13}$CO
along the PV-cut in the lower-right panel.
The high-velocity features result from the projection
of the {\it converging motion} and {\it diverging motion}.
The two dark vertical emission features correspond to
the two colliding clumps.
{\bf Bottom row:}
Channel maps for the {\tt rot30} RT-model for C$^{18}$O (left)
and $^{13}$CO (right). In the left panel,
the dashed lines mark the longitudinal PV-cuts.
{\it PV-cut1} is along the filament spine and
{\it PV-cut2} is two arcmin away from the filament.
In the right panel, the vertical line shows the transverse PV-cut for
the PV-diagram in the top-right panel.
\label{fig:transpv}}
\end{figure*}

To generate synthetic molecular line cubes,
we use the \textsc{RADMC-3D} code \citep{2012ascl.soft02015D},
following the standard line radiative transfer (RT) procedure.
Specifically, we use the LTE mode with a temperature of 15 K
which is the temperature adopted in the isothermal simulation.
We model $^{13}$CO(1-0) and C$^{18}$O(1-0) line emission,
assuming the standard ISM abundances \citep{1994ARA&A..32..191W}.
We adopt a RT cell size of 0.0039 pc, which is $\sim2\arcsec$
at the adopted distance of 400 pc and half the simulation cell size.
To match the beam size of the CARMA-NRO Orion data
\citep{2018ApJS..236...25K,2021RNAAS...5...55K},
we smooth the RT cubes to have a 8\arcsec~resolution.
The line cube covers a velocity range from
-2.5 to 2.5 km s$^{-1}$ with a channel width of 0.05 km s$^{-1}$
in order to resolve the gas kinematics.
Later when we compare with
the observation, we adopt a 0.2 km s$^{-1}$ channel width and
add a Gaussian noise of 0.5 K to
match the CARMA-NRO Orion data.
Following K21, we focus on the timestep at $t=0.6$ Myr.

Figure \ref{fig:transpv} upper-left panel shows the 
orientation used for the RT modeling, where
we follow the convention in the RADMC-3D \href{https://www.ita.uni-heidelberg.de/~dullemond/software/radmc-3d/manual_radmc3d/imagesspectra.html}{documentation}
(see their Figs 9 and 10).
Since we focus on the kinematic 
feature of the {\it converging motion} and {\it diverging motion}, 
we fix the filament to be horizontal in the synthetic image
and vary the inclination of the collision midplane
(the angle between the line-of-sight and the midplane).
Effectively, we, as the observer, rotate azimuthally
around the filament while keeping the filament horizontal in our field of view.
Specifically, under the convention of RADMC-3D,
we fix $\phi_{\rm obs}=-90^\circ$
and vary the inclination $i_{\rm obs}$ between $0^\circ-90^\circ$.
For example, with $i_{\rm obs}=90^\circ$, $y_{\rm image}$ aligns
with $z_{\rm sim}$ and $x_{\rm image}$ aligns with $y_{\rm sim}$.
Since the filament forms along $y_{\rm sim}$ ,
it aligns with $x_{\rm image}$ in the RT image,
i.e., a horizontal filament in the synthetic image.
With this orientation, the image line-of-sight
is along the negative $x_{\rm sim}$ . 

Then, if we reduce the inclination angle $i_{\rm obs}$,
the filament remains horizontal while the projected midplane changes.
Figure \ref{fig:transpv} top-left panel
shows the case with $i_{\rm obs}=30^\circ$.
Hereafter, this RT setup is referred to as the {\tt rot30} RT-model.
Its line-of-sight direction and plane-of-the-sky projection are illustrated
as purple lines and arrows in Figures \ref{fig:model} and \ref{fig:transpv}.
Later, we will also show results with $i_{\rm obs}=80^\circ$,
which is referred to as the {\tt rot80} RT-model.
The component along the line-of-sight for gas with 
{\it converging motion} or {\it diverging motion} will have higher speed
in {\tt rot30} compared to that in {\tt rot80}.

\section{Results and Analysis}\label{sec:results}

\subsection{Model Transverse PV-diagram}\label{subsec:transpv}

From Figure \ref{fig:transpv} we can see that 
the {\it converging motion} and {\it diverging motion}
should exhibit a blueshift, redshift, blueshift, redshift 
(hereafter BRBR) pattern in molecular line emission. 
In a transverse PV-diagram,
this pattern presents a special feature, which
is shown in the upper-right panel of Figure \ref{fig:transpv}
for the case of {\tt rot30}. One can see the BRBR pattern
as four relatively high-velocity features in the PV-diagram.
The two corresponding to the {\it converging motion} are
closer to the filament while the two for the {\it diverging motion}
are farther from the filament. Example emission features
in channel maps are shown in bottom panels of Figure \ref{fig:transpv}
for both C$^{18}$O and $^{13}$CO.
At -1 km s$^{-1}$, one can see the blueshift {\it diverging motion}
and the blueshift {\it converging motion} on two sides of the filament.
The former is moving away while the latter is merging into the filament.
At 1 km s$^{-1}$, the situation is similar, but reverted in redshift.

\begin{figure}[htb!]
\centering
\includegraphics[width=1.\columnwidth]{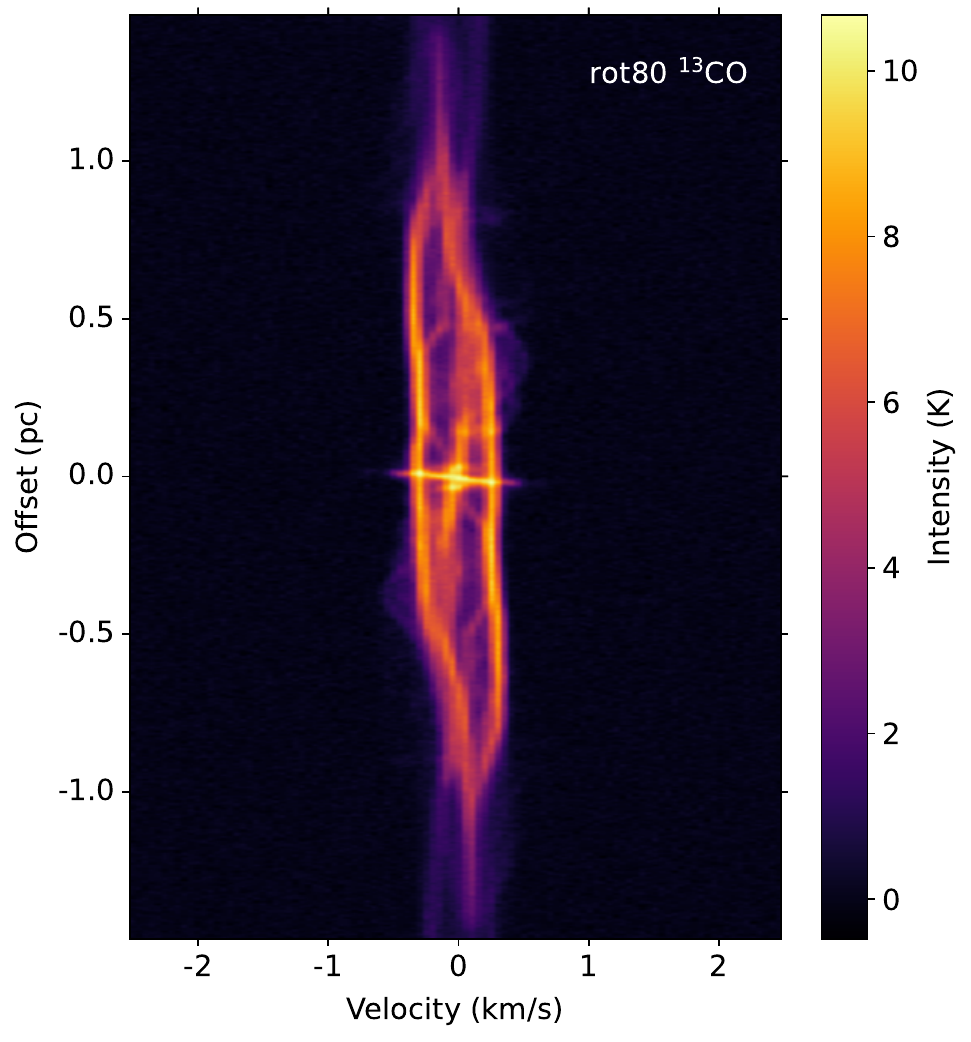}
\caption{
Transverse PV-diagram for $^{13}$CO
for the {\tt rot80} RT-model.
Due to the projection effect, no high-velocity features
similar to those in Figure \ref{fig:transpv} top-right panel
are present. The two clumps are still visible as the two
vertical emission features.
\label{fig:transpv80}}
\end{figure}

The symmetric BRBR pattern in principle
should be useful in distinguishing CMR from other models.
However, the symmetric kinematic pattern comes from the 
symmetry in the initial condition. In \citet{2023ApJS..265...58K},
we have carried out a parameter exploration and showed that
the symmetry is broken under several conditions.
On the other hand, gas in the two types of motion, while
having elevated volume density, has much lower column density
compared to the main filament. Therefore, its emission can be much weaker,
especially at large inclination ($i_{\rm obs}$),
in which case the high-velocity gas may be invisible in the transverse PV-diagram.
Figure \ref{fig:transpv80} shows an example for the {\tt rot80} RT-model.
In this case, the collision midplane is almost perpendicular to
the line-of-sight. As a result, the line-of-sight projection
of the {\it converging motion} and {\it diverging motion}
is negligible. One can see in Figure \ref{fig:transpv80},
even for the brighter $^{13}$CO line, the BRBR pattern is unnoticeable.
Additionally, the two types of motion are outside the main filament,
meaning that they are easily confused with the environment,
making it difficult to unambiguously confirm the BRBR pattern.

Longitudinal PV-diagrams along the main axis
of the filament are not impacted by these issues in principle.
As will be shown, 
these PV-diagrams reveal distinctive kinematic 
features arising from the {\it converging motion} in the 
collision midplane.

\subsection{Model Longitudinal PV-diagram}\label{sec:modelpv}

\begin{figure*}[htb!]
\centering
\includegraphics[width=0.95\textwidth]{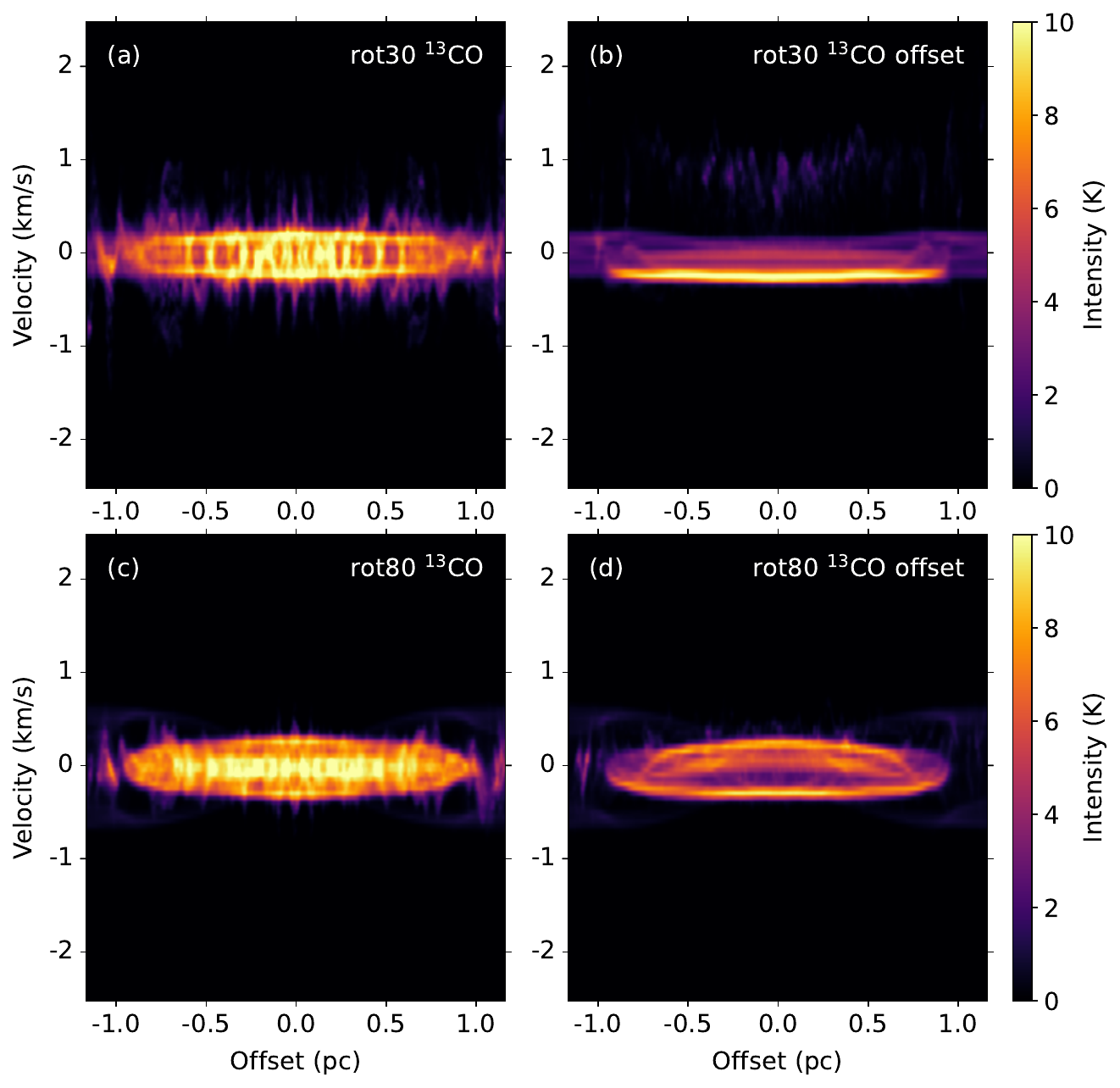}
\caption{
Model PV-diagrams for $^{13}$CO(1-0) 
along {\it PV-cut1} (left column) and {\it PV-cut2} (right column).
The top row shows results for RT-model {\tt rot30}
and the bottom row is for RT-model {\tt rot80}.
The PV-cuts are shown in Figure \ref{fig:transpv}.
\label{fig:modelPV}
}
\end{figure*}

The left column of Figure \ref{fig:modelPV} 
shows the longitudinal PV-diagram along the filament spine.
Here we only show results from the $^{13}$CO line.
The C$^{18}$O PV-diagrams show very similar structures,
except with a lower intensity.
In the longitudinal PV-diagrams,
a ``zigzag'' pattern is clearly noticeable,
indicating a velocity oscillation.
The two horizontal emission features across the oscillation 
correspond to the colliding clumps, similar to those vertical emission features seen
in the transverse PV-diagrams (Figures \ref{fig:transpv} and \ref{fig:transpv80}).
The oscillation is seen consistently along the filament.
Its spatial distribution shows a periodic characteristic
but without a fixed period, suggesting a stochastic
nature of the kinematics. The projected velocity spread is 
roughly from -0.5 to 0.5 km s$^{-1}$, which should be
observable by modern millimeter telescopes.
For the {\tt rot80} RT-model, the oscillation is still present,
although the velocity spread does not extend significantly
outside the clump velocities.

The internal kinematics of the filament is arguably 
determined by the {\it magnetic transport} defined in \S\ref{subsec:sim},
especially since the filament is at the stagnation point
of the {\it converging motion} in the collision midplane
(recall from Figure \ref{fig:model} that dense gas is 
transported to the central axis in a symmetric way).
The converging gas is clumpy,
which is readily seen in channel maps shown in Figure \ref{fig:transpv} 
and in the volume rendering in lower-left panel of
Figure \ref{fig:model}. 
In PV-diagrams, this clumpy gas appears
similar to those oscillating features, as shown in
Figure \ref{fig:modelPV}(b).
Here, we show the offset longitudinal PV-diagrams
along {\it PV-cut2} (Figure \ref{fig:transpv}).
They show kinematic features for gas in one side of
the {\it converging motion}. In panel (b), 
where we show the {\tt rot30} RT-model,
one can see the clumpy redshift emission 
at $\sim1$ km s$^{-1}$ with a velocity spread of
$\sim1$ km s$^{-1}$ (0.5-1.5 km s$^{-1}$). 
This clumpy feature, which is unnoticeable in C$^{18}$O,
comes from gas pieces that are dense enough to have
noticeable line flux. 
As these dense gas pieces are shot into the filament from two sides
by field-loops asynchronously, they trigger
the velocity oscillations seen in the longitudinal PV-diagrams. 
The asynchronous field-loops are readily seen in
Figure 2 of \citet{2023ApJS..265...58K}.
In that figure, while some field-loops already
close in on the filament as toroidal fields,
others just emerge on two sides of the {\it pancake}.
The gas inside the filament is confined by
the helical/toroidal field. Its motion also
contributes to the overall kinematics.
Because the emergence of the dense gas
and the {\it magnetic transport} continue as long as the clump
collision persists, the filament is constantly stirred
by the {\it converging motion}, maintaining the 
velocity oscillation.

On the other hand, the clumpy feature is not seen in the {\tt rot80}
offset PV-diagrams shown in Figure \ref{fig:modelPV}(d),
which is simply due to  projection effects. 
In this projection,
the collision midplane is almost perpendicular to the 
line-of-sight. As a result, the line-of-sight 
velocity component of the {\it converging motion}
(as well as the {\it diverging motion}) is negligible.
In other words, the {\it converging motion} feature
is only detectable in observation with proper projections.
Moreover, the feature can be blended with nearby structures, 
which is a problem with the transverse PV-diagram
as discussed in \S\ref{subsec:transpv}.
However, the zigzag pattern is still visible in
the {\tt rot80} filament longitudinal PV-diagram,
which is why a PV-diagram along the filament's spine is 
probably the best choice for testing CMR.

\begin{figure*}[htb!]
\centering
\includegraphics[width=0.98\textwidth]{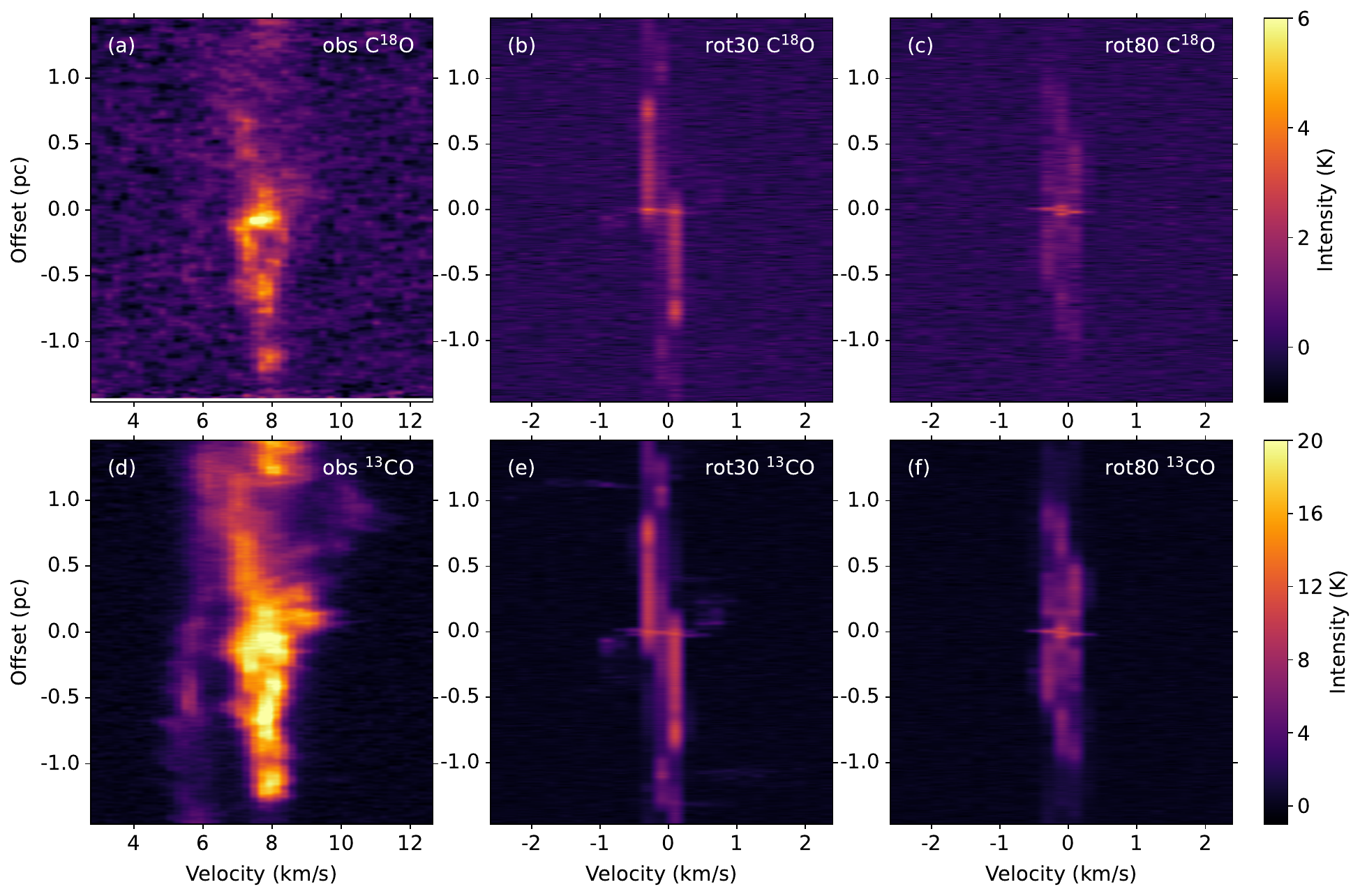}
\caption{
{\bf Left column:} Observed transverse PV-diagrams 
for C$^{18}$O (top) and $^{13}$CO (bottom).
The transverse PV-cut is from (209.8887\arcdeg, -19.3242\arcdeg)
to (210.0169\arcdeg, -19.7239\arcdeg) in Galactic coordinates.
Offset 0 is roughly at the location of the Stick filament.
{\bf Middle column:} RT-model {\tt rot30} transverse PV-diagrams 
for C$^{18}$O (top) and $^{13}$CO (bottom).
{\bf Right column:} RT-model {\tt rot80} transverse PV-diagrams 
for C$^{18}$O (top) and $^{13}$CO (bottom).
C$^{18}$O images share the same color scale 
and $^{13}$CO images share the same color scale.
The line cubes are smoothed to a velocity resolution of
0.2 km s$^{-1}$. 
\label{fig:comparetrans}
}
\end{figure*}

A deeper understanding of the clumpy feature in the
{\it converging motion} probably requires
a dedicated study under the context of plasma physics. 
As a first attempt,
a plausible explanation is that the feature is caused by
the stochastic behavior of the plasmoid in different layers
in the collision midplane.
As shown in Figure \ref{fig:model} and the associated movie, 
at any given time, the dense gas pieces in the {\it converging motion}
are scattered along the $y_{\rm sim}$-direction.
The scattering can be understood if we break the 
midplane into multiple $x_{\rm sim}$-$z_{\rm sim}$ slices.
In each slice, magnetic reconnection triggers the formation of
multiple plasmoids that are randomly distributed along
the field-reversal interface \citep{2007PhPl...14j0703L}.
They can
then move and merge with other plasmoids in the $x_{\rm sim}$-$z_{\rm sim}$ plane
\citep{2012A&A...541A..86K} in a stochastic manner.
The right panel of Figure 25 in K21 shows
an example of the process.
This behavior is why the converging and diverging gas
in channel maps (Figure \ref{fig:transpv} lower panels)
appear so clumpy. The dense gas pieces in 
the collision midplane 
are an aggregation of the plasmoids
in different layers that are accelerated
by the field-loops. Since the formation and merging of
the plasmoids turn out to be chaotic, the {\it converging motion}
in CMR becomes clumpy and shows the clumpy kinematic feature.

In classical resistive MHD, the plasmoid instability grows
monotonically with the Lundquist number:
\begin{equation}\label{equ:lund}
S_L=\frac{Lv_A}{\eta},
\end{equation}
where L is the current sheet length,
$v_A=B/\sqrt{4\pi\rho}$ is the upstream Alfv\'en speed,
and $\eta$ is the Ohmic resistivity.
The instability develops when $S_L$ reaches a high value
\citep[$\sim10^4$, see, e.g., ][]{tajima_shibata,2017ApJ...850..142C}.
In our model, $S_L\sim 10^4$, matching
the quoted value from literature. 
However, it is possible $S_L\lesssim10^3$ under certain conditions.
In that case, the {\it pancake} may not be so clumpy.
Then, it is unclear if the oscillating pattern remains.
This uncertainty will be investigated in the future.

\subsection{Comparison with Stick observations}\label{sec:obspv}

\subsubsection{Transverse PV-diagrams}\label{subsubsec:transpv}

The BRBR pattern was not obvious in the observed
PV-diagram in K21 Figure 5 as it was not extended far enough
from the filament to include the {\it diverging motion}.
We re-examine the PV-diagram in both C$^{18}$O and $^{13}$CO
cubes from the CARMA-NRO Orion data
by extending the PV-cut to a length of 2.9 pc
that match the RT-model PV-cut length (Figure \ref{fig:transpv}).
Figure \ref{fig:comparetrans} shows the results.
The observed PV-diagrams are shown in the left column.
The main filament is roughly at velocity 8 km s$^{-1}$ and offset 0.
If we focus on the {\tt rot30} RT-model first by comparing
panel (a) and (b), the BRBR pattern is not clear in C$^{18}$O,
although there is a hint of high-velocity feature
around offset 0.2 pc and velocity 9 km s$^{-1}$ in observation.
Here, ``high-velocity feature'' refers to those features
that are separated from the main filament velocity
in the transverse PV-diagram.
In $^{13}$CO observation in panel (d), 
there are high-velocity structures at
6 km s$^{-1}$ and 10 km s$^{-1}$.
But they do not exactly
match the RT-model PV-diagram in panel (e). 
If we focus on the {\tt rot80} RT-model in panels (c) and (f),
the main filament appears to match the observation better,
especially for features at negative offsets.
However, as with Figure \ref{fig:transpv80}, 
there is no noticeable high-velocity feature in {\tt rot80}.

\begin{figure*}[htb!]
\centering
\includegraphics[width=0.98\textwidth]{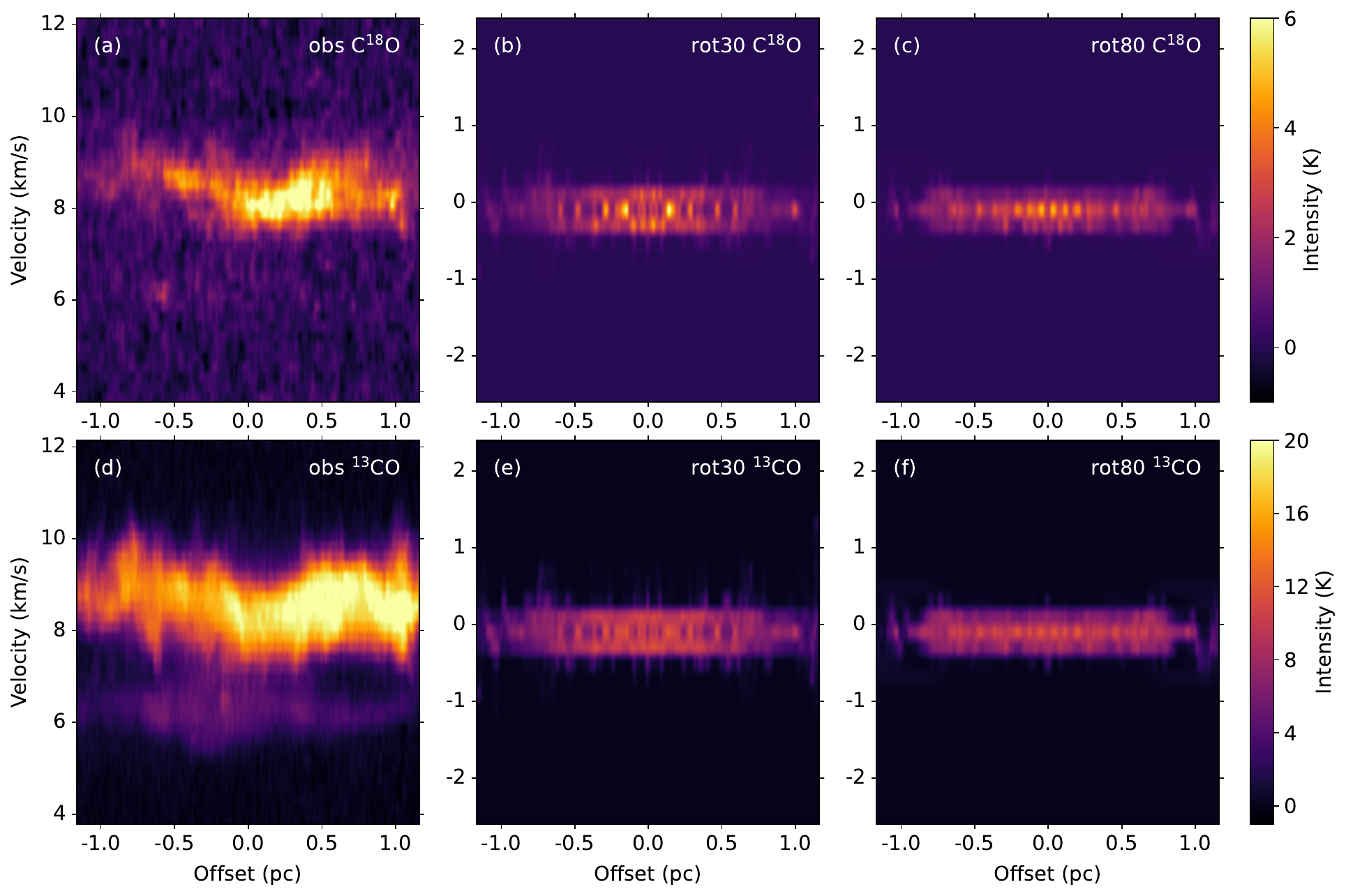}
\caption{
Comparison of longitudinal PV-diagrams between observation and simulation.
{\bf Left column:} Longitudinal PV-diagrams along the Stick filament with
the CARMA-NRO Orion data for C$^{18}$O (top) and $^{13}$CO (bottom).
The centroid velocity of the Stick filament is at $\sim8.3$ km s$^{-1}$.
{\bf Middle column:} PV-diagrams for RT-model {\tt rot30} for
C$^{18}$O (top) and $^{13}$CO (bottom).
{\bf Right column:} PV-diagrams for RT-model {\tt rot80} for 
C$^{18}$O (top) and $^{13}$CO (bottom).
The RT-model data now has a 0.2 km s$^{-1}$ channel width
which is similar to the 0.22 km s$^{-1}$ channel width in observation.
\label{fig:comparelongi}}
\end{figure*}

So there are two possibilities. First, those hints of
high-velocity features belong to the filament system, 
and the {\tt rot30} RT-model better reproduces the observation.
Alternatively, the filament system is better reproduced by
the {\tt rot80} RT-model, while those high-velocity features
in $^{13}$CO observation belong to other line-of-sight gas structures.
Note, the observed line intensity is about a factor of 2
higher than the RT-models. Either the simulation needs to be adjusted
(e.g., by increasing the initial gas density) or the observation
overestimates the line intensity. In the CARMA-NRO Orion data
paper \citep{2018ApJS..236...25K}, there was an intensity mismatch
between the single-dish data and the interferometer data.
After an extensive examination, they scaled the single-dish
data up by a factor of 1.6 to match the interferometer data.

Another mismatch between the RT-model and the observation is
that the velocity spread in the observed PV-diagrams is
roughly wider than the RT-model by a factor of 2.
Here, the velocity spread refers to the separation
between the two vertical emission features. As discussed earlier,
they correspond to the colliding clumps. In principle,
the mismatch can be alleviated by increasing the initial colliding speed
or the shear velocity in the simulation by the corresponding factor.
The projection effect can be another cause.
Here, the key point of the comparison in Figure \ref{fig:comparetrans}
is to match the characteristics 
between the simulation and the Stick.
We did not tweak model parameters to exactly reproduce the velocity spread.
Note, the linewidth of the filament, taken at offset 0, 
is different from the velocity spread. While the test simulation in
\S\ref{subsec:bpad} and Figure \ref{fig:bpad}
shows that a factor of 2 larger initial colliding
speed results in a larger oscillating velocity spread,
other studies \citep[e.g.,][]{2023MNRAS.522.3890P} have shown that
a faster collision does not necessarily translate into a broader
filament linewidth. We also point out that the filament emission
in our RT-models appears sharper than the observations.
The sharpness could be unrealistic as non-ideal MHD effects
may be important at smaller scales, so that our simple
resistive-MHD framework is too idealized.

Besides the reasoning in \S\ref{subsec:transpv} for the possibility
of not finding BRBR, another possibility is that gas in the 
{\it converging motion} and the {\it diverging motion}
is not molecular. Molecular gas formation
is not included in the current isothermal, single-fluid
simulation. The CO gas emission is based on the simple conversion
by a constant abundance. In realistic ISM conditions,
gas in the two types of motion may well be atomic, which
only turns into molecular as the density increases in the main filament.
And even then it may take a while until CO gas becomes visible
\citep{2019MNRAS.486.4622C,2024ApJ...975...97K}.
The \textsc{Arepo} simulation of colliding atomic gas
in \S\ref{subsec:isf} gives an example.
Transverse PV-diagrams of CO only show the central filament
without the two colliding clouds or the high-velocity converging gas.
For detecting BRBR, perhaps (a combination of) other tracers
can be used jointly, including HI and H$_2$.

\subsubsection{Longitudinal PV-diagrams}\label{subsubsec:longipv}

Figure \ref{fig:comparelongi} shows the comparison of longitudinal PV-diagrams.
The left column shows PV-diagrams along the
Stick filament in C$^{18}$O(1-0) and $^{13}$CO(1-0).
The data cubes are from the 
CARMA-NRO Orion Survey \citep{2018ApJS..236...25K,2021RNAAS...5...55K}.
While there is a hint of velocity oscillation in the C$^{18}$O PV-diagram,
the $^{13}$CO PV-diagram shows the oscillation better.
The middle column and the right column show PV-diagrams
for RT-model {\tt rot30} and RT-model {\tt rot80}, respectively.
Compared to Figure \ref{fig:modelPV}, the RT-model data are
now with a channel width of 0.2 km s$^{-1}$ channels,
similar to that of the observations.
The centroid velocity of the RT-model filament is at 0.
The velocity oscillation in the RT-model is still visible
even with the lower velocity resolution. 
Qualitatively, both the observation and the RT-model show
two characteristics. First, there are relatively brighter blobs
around the centroid velocity. Second, the velocity oscillation
stretches out at higher velocities across those blobs.
One major difference is that the spatial oscillating frequency
in observation appears to be higher than that in simulations.

\section{Discussion}\label{sec:discus}

\subsection{Collision with an initial separation}\label{subsec:bpad}

\begin{figure*}[htb!]
\centering
\includegraphics[width=0.98\textwidth]{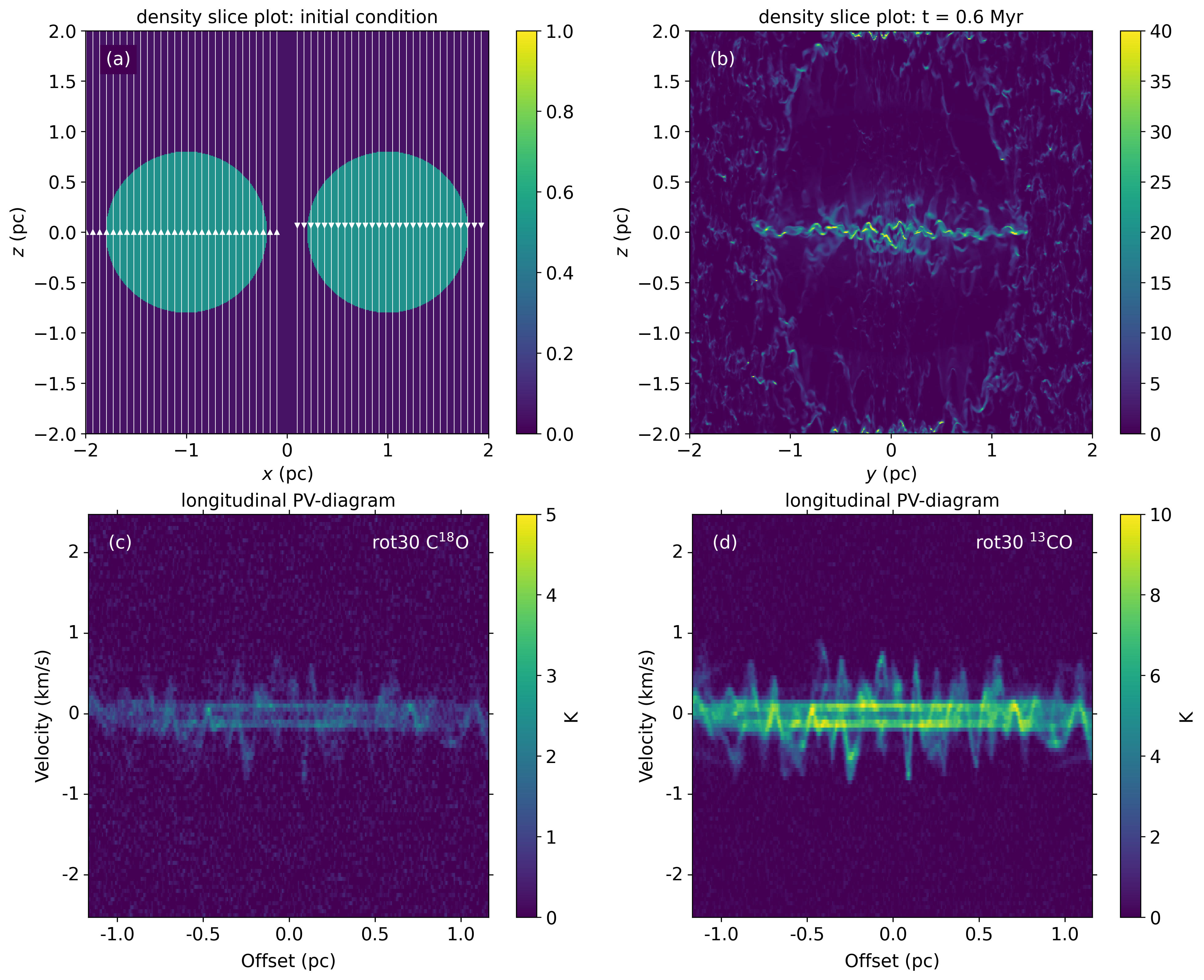}
\caption{
{\bf (a):} Density slice plot for the initial condition for 
simulation \mbox{test\_Bpad}.
The colliding clumps are separated by 0.4 pc 
and the reverse fields are separated by 0.2 pc. 
{\bf (b):} Density slice plot for the collision midplane
at t=0.6 Myr. Again, it shows that a filament forms instead of a pancake.
Compared to the fiducial case, the filament here appears more wiggly.
{\bf (c):} Longitudinal PV-diagram for C$^{18}$O(1-0) for RT-model {\tt rot30}
for the new simulation at t=0.6 Myr.
{\bf (d):} Longitudinal PV-diagram for $^{13}$CO(1-0) for RT-model {\tt rot30}
for the new simulation at t=0.6 Myr.
\label{fig:bpad}}
\end{figure*}

The setup of the fiducial simulation
has the two clumps and the reverse field
contacting at the $x_{\rm sim}=0$ interface.
However, it is not a necessary condition for CMR.
In Figure \ref{fig:bpad}, we show a test simulation (hereafter
\mbox{test\_Bpad}) with the clumps and the reverse field 
separated initially. For simplicity, we do not expand the computation
domain as \mbox{MRCOLe2}. So the boundaries are limited at
$\pm2$ pc. Compared to the fiducial simulation, the initial reverse
field is separated by 0.2 pc, and the initial clumps are
separated by 0.4 pc. Due to the separation and the limited
computation domain, we shrink the clump radius from 0.9 pc to 
0.8 pc and increase the colliding speed by a factor of 2
to shorten the time before collision. We also
remove the shear speed to have a head-on collision.

As shown in Figure \ref{fig:bpad} top row, an initial separation
between the magnetic field and the clumps does not prevent
the onset of CMR or the subsequent formation of a filament.
A similar study was carried out by \citet{2022ApJ...933...40K},
who considered initially separated clouds with a smooth
field reversal in a bisymmetric spiral magnetic field configuration.
In realistic ISM environments, large-scale reversed
magnetic fields can be brought into contact by colliding
flows or expanding bubbles. Simulating
such conditions is beyond the scope of this paper.
But it is an interesting direction for future work.

We apply the same radiative transfer procedure from
\S\ref{subsec:so} to the test simulation \mbox{test\_Bpad}
and make longitudinal PV-diagrams for $^{13}$CO and C$^{18}$O.
Figure \ref{fig:bpad} bottom row shows the results.
One can see that the filament again shows the oscillating
kinematics along its spine. Since we have increased the
collision speed by a factor of 2, the oscillating velocity
spread is widened and comparable to those seen in the Stick
(Figure \ref{fig:comparelongi}). This result suggests that
the oscillation is a universal kinematic pattern
in CMR.

\subsection{The integral-shaped filament}\label{subsec:isf}

\begin{figure*}[htb!]
\centering
\includegraphics[width=0.98\textwidth]{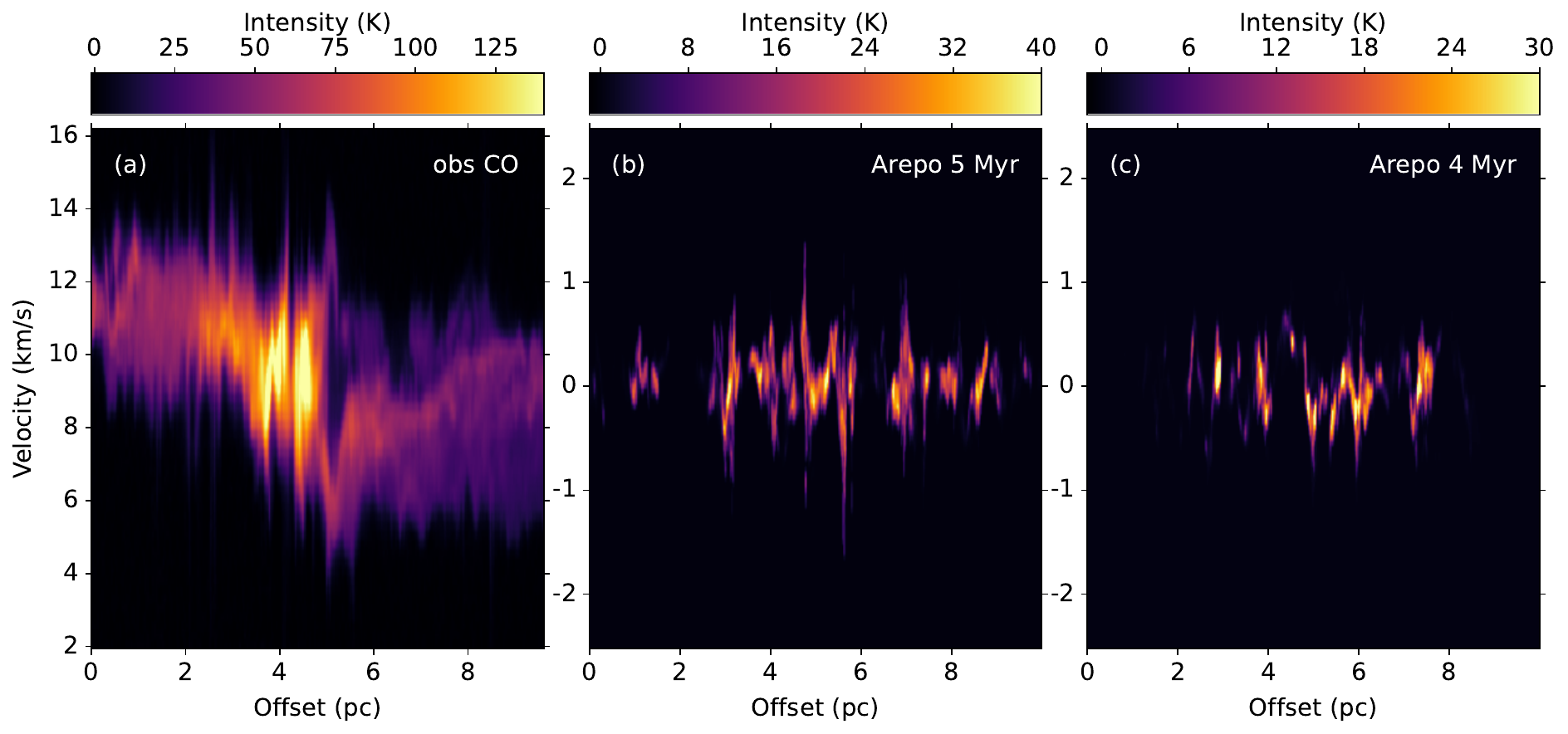}
\caption{
{\bf (a):} Longitudinal PV-diagrams along the integral-shaped filament
in CO(1-0) from the CARMA-NRO Orion data.
See \citet{2018ApJS..236...25K} Figs 20-22 for another view
of the PV-diagram, including CO(1-0), $^{13}$CO(1-0), and C$^{18}$O(1-0).
All three isotopologues show the oscillating kinematics.
{\bf (b):} Longitudinal PV-diagram for CO(1-0) for the simulation 
from \citet{2024ApJ...975...97K}. Here the panel shows the
result from snapshot at 5 Myr.
{\bf (c):} Longitudinal PV-diagrams for the same simulation
in panel (b) but from snapshot at 4 Myr.
\label{fig:isfarepo}}
\end{figure*}

The oscillating pattern in the Stick filament is reminiscent of
the large-scale velocity oscillation in the integral-shaped filament
in Orion A seen in \citet{2018ApJS..236...25K}.
With the CARMA-Orion NRO data,
they made longitudinal PV-diagrams
along the spine of the Orion A integral-shaped filament
in CO, $^{13}$CO, and C$^{18}$O. All three PV-diagrams
showed a clear velocity oscillation over the $\sim10$ pc length.
\citet{2019MNRAS.489.4771G,2019MNRAS.487.1259L} also saw
similar kinematics in Orion A and other clouds.
In fact, this feature is another motivation for this paper.
If the oscillation is a characteristic of CMR,
then it provides a possible explanation for
the PV-diagrams for Orion A. However, the scale
of the observed filament is much larger than that in the fiducial simulation.
To examine whether CMR at larger scale also possesses
the oscillating characteristic, one needs to simulate
the process at a larger scale.
\citet{2024ApJ...975...97K} have used
\textsc{Arepo} \citep{2010MNRAS.401..791S}
to simulate the CMR process at a scale 10 times larger
than \mbox{MRCOL}. They started with atomic gas collision,
and the simulation produced a molecular filament 
with a length of $\sim20$ pc at 5 Myr.
Since their filament has a similar scale as
the Orion A filament, we examine their filament
kinematics for the oscillation and compare them with the Orion A observations.
Other details of the simulation and the synthetic observations
can be found in \citet{2024ApJ...975...97K}.

Figure \ref{fig:isfarepo} shows the comparison.
Panel (a) shows the longitudinal PV-diagram along
the integral-shaped filament with the CO(1-0) data.
Unlike \citet{2018ApJS..236...25K} who defined a
30-point polyline as the PV-cut, we simply adopt a
4-point polyline that roughly follows the main filament.
As shown in panel (a), the different choice of the PV-cut
still shows the oscillation well. The four points are
(208.445\arcdeg, -19.295\arcdeg),
(208.771\arcdeg, -19.209\arcdeg),
(209.441\arcdeg, -19.676\arcdeg),
(209.725\arcdeg, -19.627\arcdeg) in Galactic coordinates.
Panel (b) and panel (c) show the longitudinal PV-diagrams
along the simulated filament from \citet{2024ApJ...975...97K}.
We select the central 10 pc segment of the filament
to match the length of the integral-shaped filament.
Two snapshots at 5 Myr and 4 Myr
are shown in the two panels, respectively. 
They are selected because there is significant molecule formation
at these times. 

Figure \ref{fig:isfarepo}(a) shows a similar
velocity oscillating pattern as those seen in
\citet{2018ApJS..236...25K} Figs. 20-22.
Figure \ref{fig:isfarepo} panels (b) and (c) clearly show
a velocity oscillation in the simulation at 10 pc scales.
Interestingly, we do not see the two horizontal emission
features across the oscillation as in the \textsc{Athena++}
simulations. This is simply because the new simulation
started with atomic gas, so the colliding clouds are
not visible in the PV-diagrams for CO isotopologues.
Since both the \textsc{Athena++} and the \textsc{Arepo}
simulations produce the oscillating kinematics,
we argue that the oscillating pattern is an intrinsic
characteristic of CMR, which provides a useful test with observations.
Since the pattern shows up at scales of 2 pc and 20 pc,
it is plausible to hypothesize that the emergence of this characteristic 
is scale-free, although the oscillating frequency may have limits.
Answering this question requires simulations at scales
across orders of magnitude, which is out of the scope of this paper.
The line intensity of the observation is a factor of
$\sim3$ times higher than the \textsc{Arepo} simulation.
Besides the scale-up factor in the CARMA-NRO-Orion data
(see discussion in \S\ref{subsubsec:transpv}),
it is possible that the simulated filament would become
brighter in CO if their simulation continued longer after t=5 Myr.
The velocity spread in the observation is about a factor of 2 wider
than the simulation, which could be remedied if the
initial colliding speed is increased by a factor of two
(see the example in \S\ref{subsec:bpad}).
Again, these two discrepancies could
result from the fact that the \textsc{Arepo} simulation
was not designed specifically for Orion A. 
The key point here is that the characteristic from CMR
gives an explanation to the oscillating pattern
seen in observation \citep{2018ApJS..236...25K}.

Observation has suggested that the integral-shaped filament
is wrapped by a helical field \citep{2011ApJ...741..112P}.
The Orion A cloud is between a large-scale field reversal
\citep{1997ApJS..111..245H,2018A&A...614A.100T}.
It is reasonable to hypothesize that Orion A
formed via CMR on a larger scale. A plausible scenario
is that the large-scale field reversal represents
the initial condition, and a gas collision
(potentially powered by expanding bubbles) 
with this reverse field results in a filamentary cloud. 
As a natural result of CMR, the filament is
wrapped by helical fields.
Exactly reproducing the oscillation
and the Orion A formation will be pursued in the future.
Yet, we believe the velocity oscillation is a
useful characteristic for testing CMR.
It is possible that other simulations
or models also show similar features and can explain the
Stick and the Orion A formation. Comparison with these
models would not only clarify the formation history of 
the Orion A cloud, but also have intriguing implications
on the formation and evolution of molecular clouds in general.

\section{Summary}\label{sec:sum}

To summarize, we have shown kinematic features of the CMR mechanism
to provide testable features with observations.
These features primarily emerge from
the collision midplane in which magnetic reconnection
and {\it magnetic transport} happen. In particular,
due to the nature of CMR, the collision midplane has a 
{\it converging motion} and a {\it diverging motion}.
The two types of motion exhibit a special 
blueshift, redshift, blueshift, redshift (BRBR)
velocity pattern if we observe the midplane
with an inclination angle. However, the BRBR
pattern can be unnoticeable if the inclination is high.
The pattern can also be ambiguous in the complex interstellar
medium due to emission blending. 

The longitudinal PV-diagram along the CMR-filament spine
shows a prominent velocity oscillation pattern,
which is barely impacted by the inclination.
The pattern persists even when the collision
midplane is almost perpendicular to the 
line-of-sight. With such high inclination,
the BRBR pattern would be nearly unnoticeable
due to projection effect. The oscillation pattern
is seen in both the Stick filament and
the integral-shaped filament PV-diagrams,
although the exact matching of the
oscillation frequency and the velocity spread
need further investigation. 
Since the PV-diagram is along the filament 
spine, the oscillation pattern is less 
impacted by confusion with nearby gas emission.
Therefore, we believe the oscillation pattern
in the spinal PV-diagram provides a useful kinematic
feature for testing CMR with observations.

\begin{acknowledgments}

We thank an anonymous referee for the constructive report.
SK acknowledges fruitful discussion with Felix Priestley and Tianwei Zhang.
An allocation of computer time from the UA Research Computing High Performance Computing (HPC) at the University of Arizona is gratefully acknowledged.
RSK acknowledges financial support from the ERC via Synergy Grant ``ECOGAL'' (project ID 855130) and from DFG and ANR for project ``STARCLUSTERS'' (funding ID KL 1358/22-1). 
VOO is supported by Cluster of Excellence ``Our Dynamic Universe'', funded by the Deutsche Forschungsgemeinschaft (DFG, German Research Foundation) under Germany's Excellence Strategy EXC 3037 – 533607693 – Unser dynamisches Universum.
H.G.A. is supported in part by NSF grant AST-2407116.
\end{acknowledgments}

\software{Astropy \citep{Astropy-Collaboration13}, pvextractor \citep{2016ascl.soft08010G}, Numpy \citep{numpy}, Matplotlib \citep{matplotlib}, SAOImageDS9 \citep{2003ASPC..295..489J}, CARTA \citep{carta}, VisIt \citep{HPV:VisIt}.}

\facility{UA HPC};

\bibliography{ms}
\bibliographystyle{aasjournal}

\end{document}